\documentstyle[multicol,prl,aps,epsfig]{revtex}
\begin{document}
\draft
\title{Anomalous Chiral Luttinger Liquid Behavior of Diluted Fractionally Charged Quasiparticles }
\author{Y.~C.~Chung, M.~Heiblum, Y.~Oreg, ~V.~Umansky, and D.~Mahalu}
\address{Braun Center for Submicron Research, Department of Condensed
 Matter Physics, Weizmann Institute of Science, Rehovot 76100, Israel}
\date{\today}
\maketitle

\begin{abstract}
Fractionally charged quasiparticles in \textit{edge states}, are
expected to condense to a \textit{chiral Luttinger liquid} (CLL).
We studied their condensation by measuring the conductance and
shot noise due to an artificial backscatterer embedded in their
path. At sufficiently low temperatures backscattering events were
found to be strongly correlated, producing a highly non-linear
current-voltage characteristic and a non-classical shot noise -
both are expected in a CLL.  When, however, the impinging beam of
quasiparticles was made dilute, either artificially via an
additional weak backscatterer or by increasing the temperature,
the resultant outgoing noise was classical, indicating the
scattering of independent quasiparticles.  Here, we study in some
detail this surprising crossover from correlated particle behavior
to an independent behavior, as a function of beam dilution and
temperature.

\end{abstract}

 \noindent

\begin{multicols}{2}
When electrons, confined to two dimensions, are subjected to an
extremely strong magnetic field, their orbits quantize and Landau
levels are formed.  Electrons that occupy only a fraction
$\nu=l/m$ (called \textit{filling factor}, with $m$ odd and $l$
integer) of the first Landau level form the so-called Laughlin
quasiparticles. Being independent each quasiparticle carries a
charge $e/m$~\cite{r1}.  The main characteristics of that regime
are the zeros of the longitudinal conductance and the exact
plateaus of the transverse conductance $g=\nu e2/h$, with $e$ the
electron charge and $h$ Plank's constant.  This is the well-known
fractional quantum Hall effect (FQHE)~\cite{r2} .  The current,
carried by the quasiparticles, flows in narrow, one-dimensional
like, strips along the edges of the sample in quantized edge
states~\cite{r3}.  Wen predicted~\cite{r4}  that being confined to
the edge the fractionally charged quasiparticles are expected to
form a \textit{non-Fermi liquid} (FL) system, a \textit{chiral
Luttinger liquid} (CLL). The validity of the CLL model can be
tested, for example, by studying the effect of a backscattering
potential on the conductance and on the \textit{shot noise}. Such
potential induces charge density wave in the 1D channel, leading
to correlation among the scattering events - not like in a FL
where the events are independent.  Typically, even the weakest
backscattering potential is expected to quench the longitudinal
conductance at zero temperature with a non-linear I-V
characteristic that is highly temperature
sensitive~\cite{r5,r6,r7,r8}. The resultant shot noise, in turn,
is predicted to be non-classical (non-Poissonian), with a voltage
(or current) dependent scattered charge~\cite{r9,r10}.

What had been already known?  In the weak backscattering regime,
at sufficiently high temperature (greater than the characteristic
backscattering energy), correlation among scattering events is
weak with a classical-like shot noise.  Noise is proportional to
the reflected current and the quasiparticle charge~\cite{r11} , as
was demonstrated for filling factor  $\nu=1/3$ and $\nu=2/5$ by
deducing a quasiparticle charge $e/3$ and $e/5$,
respectively~\cite{r12,r13,r14}. In the strong backscattering
regime~\cite{r9,r10,r11} only \textit{electrons}, or bunched
quasiparticles, are allowed to tunnel through opaque
barriers~\cite{r15}. Contrary to that well established behavior, a
most recent experiment proved that highly dilute quasiparticles
(quasiparticles arrive \textit{one by one}) traverses an
\textit{opaque} barrier without bunching, namely, the scattered
charge is nearly $e/3$~\cite{r16} . This unexpected result cannot
be presently explained by theory.
This unexpected behavior led us
to concentrate on the transport of $e/3$ quasiparticles in very
dilute beams (10 $\sim$ 20 $\%$) or in the fully occupied regime,
and at a wide range of electron temperature ($20 \sim 120mK$).
While finding an excellent agreement with the CLL prediction for
fully occupied beams at low temperature - low energy regime, we
observed a clear transition toward an independent particle
behavior of highly dilute beams. We conclude that beam dilution
plays a qualitative similar role to that of temperature, a regime
where theory is still lacking~\cite{r17}.

Measurements were conducted at bulk conductance $g_Q=e^2/3h$
plateau (B $\sim$ 13.1T).  Two QPCs were formed in a
two-dimensional electron gas (2DEG), embedded in a GaAs-AlGaAs
heterojunction, as seen in Fig. 1a.  QPC1 was used to dilute the
quasiparticle beam and QPC2 to serve as the backscattering
potential.  A many-terminal-configuration was employed in order to
prevent multiple scatterings between the two QPCs and keep the
input and output conductance constant, $e^2/3h$ - independent of
the transmission of each QPC~\cite{r15,r16}.  The differential
conductance was measured with AC excitation of 1.5$\mu$V at 3Hz
superimposed on DC bias.  The spectral density S of the shot noise
was measured as a function of DC current at a center frequency of
1.4MHz and bandwidth of 30kHz (determined by a LC resonant
circuit; see Refs. 12-13 for more details).  Voltage fluctuations
in terminal \textbf{A} were amplified by a low noise cryogenic
amplifier followed by a spectrum analyzer, which monitored the
average square of the amplified fluctuations.

\begin{figure}
\begin{center}
\leavevmode \epsfxsize=8.5 cm  \epsfbox{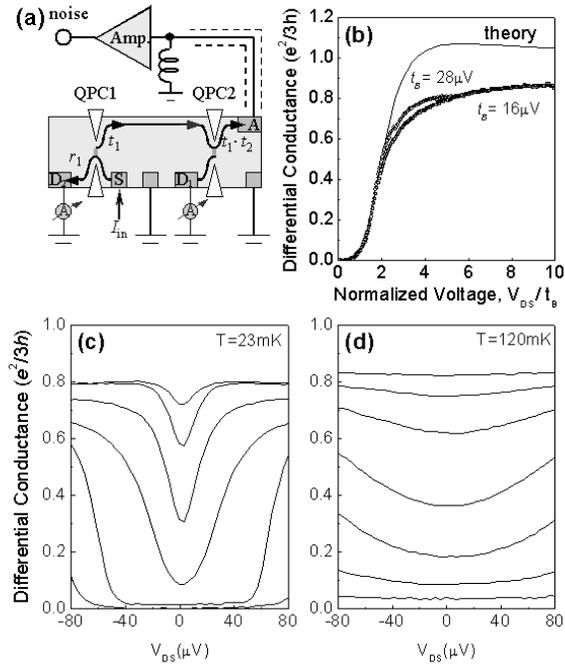}
\end{center}
\vspace{-0.5 cm} \caption{ \textbf{(a)} Schematic diagram of the
fabricated device and the measurement set up.  The device is
fabricated in a 2DEG with mobility 2 $\times$ 10$^6$
cm$^2$v$^{-1}$s$^{-1}$ and carrier density 1.1 $\times$
10$^{11}$cm$^{-2}$, at 4.2K.  Current is injected via source
\textbf{S} and is scattered by QPC1 toward QPC2. Transmission of
each QPC is measured by measuring the currents at drains D1 and
D2.  Shot noise is measured by monitoring the voltage fluctuation
at Ohmic contact \textbf{a}, after amplification by a cryogenic
amplifier with input current noise 6.7$\times$10$^{-29}$A$^2$/Hz.
The noise spectrum is filtered by a LC circuit, which is tuned to
a center frequency of 1.4MHz (with bandwidth of 30kHz).  Multiple
terminals assure constant sample conductance $e^2/3h$ and a
constant equilibrium noise. \textbf{(b)} Summary of the
differential conductance measured at 23mK for different settings
of the QPC constriction. It is plotted against an applied voltage
normalized by two effective scattering potentials, $V_{DS}/t_B$
($t_B=28\mu V$ and $t_B=16\mu V$ are measured at $V_g=-0.107$V and
-0.093V, respectively). The theoretical prediction at 0K is shown
for comparison. \textbf{(c)} Differential conductance of QPC1 as a
function of $V_{DS}$ measured at 23mK, for a few backscattering
potential strengths (gate voltage is -83mV, -75.5mV, -65.5mV,
-50.5mV, -43mV and -38.5mV, from bottom to top). \textbf{(d)}
Similar data as in C but measured at 120mK (gate voltage is 111mV,
-101mV, -91mV, -81mV, -71mV, -61mV, and -51mV, from bottom to
top).}
\end{figure}

\noindent The temperature $T$ of the electrons was determined by
measuring the equilibrium noise, $S=4k_BTg$, with $g$ the sample
conductance.  Shot noise was determined by subtracting the DC
current independent noise from the total noise signal.

Figures 1c and 1d show typical differential conductance of a
single QPC (say QPC1) as a function of the applied voltage
$V_{DS}$ at electron temperatures $T$=23mK and $T$=120mK,
respectively, for different backscattering potential strengths
(determined by the gate voltage of the QPC).  At the lower
temperature, even a relatively weak backscattering potential (with
a saturated reflection $r\sim0.3$ or transmission $t=g/g_Q \sim
0.7$), reflects almost fully the current at zero applied voltage.
The differential conductance was compared with Fendley's et al.
prediction~\cite{r9,r10} in Fig. 1b, which poses a universal
dependence on the applied voltage normalized by the so-called
\textit{impurity strength}, $t_B$, at zero temperature.  An
excellent fit with experiment is seen at the low energy regime for
$t_B$ in the range $10 \sim 40 \mu$ eV (with $k_BT \ll t_B$), as
theory requires). At high bias the differential conductance is
expected to exceed $g_Q=e^2/3h$ , however we have never observed
it in the experiments.  This can be justified if we note that in
that range short-range, non-universal, physics is dominant, making
the agreement poor. When the temperature increased to 120mK (Fig.
1d), the non-linearity weakened significantly - resembling a FL
behavior.

While independent particle scattering is stochastic, with a
classical-like shot noise, correlated particles scattering leads
to non-universal shot noise that depends on the type of the
correlation.  We recall the expression of shot noise for
independent scattering events at finite temperatures~\cite{r18} :
\begin{equation}\label{Eq1}
S=2qV_{DS} g_Q t(1-t) \left[ \coth \left( {{qV_{DS}} \over
{2k_BT}} \right) - {{2k_BT}\over{qV_{DS}}} \right]
\end{equation}
\noindent with $q$ the partitioned charge.  We plot in Fig. 2a the
differential conductance (in the inset) and shot noise both
measured at different temperatures with a relatively weak
backscattering potential (saturated $t\sim 0.8$).  At the lowest
temperature, 23mK, the differential conductance dipped near
$V_{DS}=0$ and the shot noise deviated considerably from the
independent particle behavior.  The noise is seen to increase fast
with increasing current - indicative of a high effective charge,
while later it saturates - indicative of a smaller effective
charge. The expected noise of independently scattered charges (Eq.
1) with $q=e/3$ was plotted for comparison.  As the temperature
increased the measured nonlinearity weakened and the measured shot
noise at 120mK agreed with the classical prediction.  Such, high
temperature, charge determination had been extensively employed
before in order to determine the charge of the
quasiparticles~\cite{r12,r13,r14}.

\begin{figure}
\begin{center}
\leavevmode \epsfxsize=8.5 cm  \epsfbox{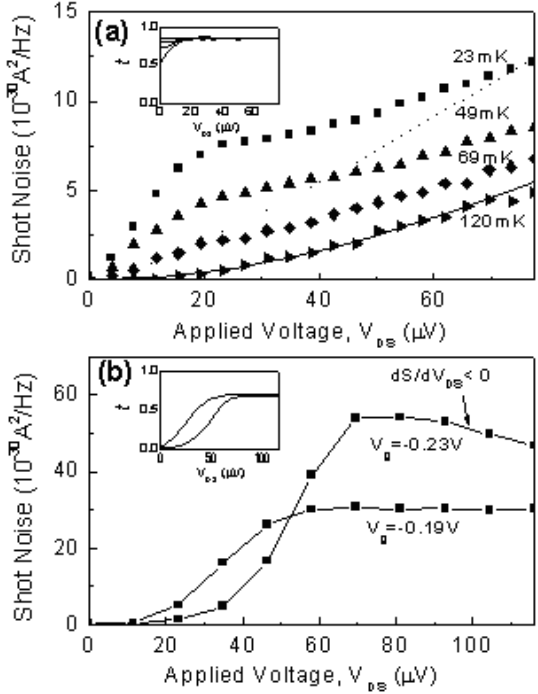}
\end{center}
\vspace{-0.5 cm} \caption{The shot noise and differential
conductance for a single QPC. \textbf{(a)} Shot noise due to a
weak backscattering potential measured at various temperatures,
23, 49, 69 and 120mK. Solid lines are the expected shot noise for
non-interacting quasiparticle with charge $e/3$. Inset: The
differential conductance of the QPC at the same temperatures. Note
the strong non-linearity at lower temperatures. A clear crossover
from interacting to non-interacting behavior is seen as the
temperature increases. This is evident at 120mK, where a linear
dependence of the noise on voltage (above some 40$\mu V$) is
observed. \textbf{(b)} Shot noise due to strong backscattering
potentials measured at 23mK. The voltage on the gates of the QPC
is -0.23V and -0.19V. Note the negative slope of the noise at high
applied voltage $(dS/dV_{DS}<0)$.  Inset: The differential
conductance of the QPC for the same gate voltages of the QPC.}
\end{figure}

To stress further the fact that scattering events of
quasiparticles at the lowest temperature are correlated we plotted
in Fig. 2b the conductance and shot noise for a stronger
backscatterer (saturated $t \sim 0.6$).  Remarkably, the noise
saturated or even changed the sign of the slope with increasing
bias to above 70$\mu$V.  In other words, adding high-energy
quasiparticles to the beam lowered the noise of the low-energy
quasiparticles. Since the backscattering strength is independent
of bias (not shown here), stochastic partitioning could never
explain such noise behavior.  This is a clear and direct
observation of the interaction among electrons in a CLL.

One can naively ask whether an artificial decrease of the average
occupation of the incoming states might weaken the correlation
among scattering events, hence rendering the scattered
quasiparticles independent - much like a temperature increase.

\begin{figure}
\begin{center}
\leavevmode \epsfxsize=8.5 cm  \epsfbox{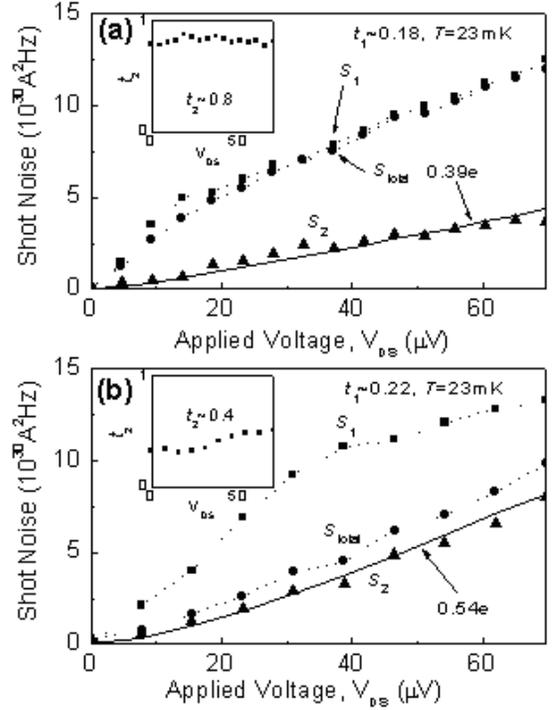}
\end{center}
\vspace{-0.5 cm} \caption{Shot noise resulting from a diluted beam
(with average occupancy $t_1 \sim 0.2$) impinges on QPC2 with
transmission $t_2$. Noise generated by QPC1, $S_1$, in squares and
the total noise measured at A, $S_{total}$, in circles. The
deduced noise generated by QPC2, $S_2$, is plotted in triangles
and is compared with shot noise expected for a binomial
partitioning of quasiparticles with charge $q_2$ (plotted in solid
line). Insets: The differential transmission of QPC2, $t_2$, of an
impinging beam of dilute quasiparticles.  The transmissions are
only slightly energy dependent.  \textbf{(a)}  $S_2$ generated by
a relatively open QPC2, $t_2 \sim 0.8$, and the deduced effective
charge $q_2=0.39e$.  \textbf{(b)} $S_2$ generated by a relatively
pinched off QPC2, $t_2 \sim 0.4$, and the deduced effective charge
$q_2=0.54e$.}
\end{figure}

\noindent This can be tested via employing the \textit{diluting
technique} first presented in Ref. 16 (shown in Fig. 1A).  The
relatively open QPC1 backscatters dilute quasiparticles with an
average occupation of each state determined by it transmission
toward QPC2, $t_1 \approx 0.1 \sim 0.2$. This beam is already
noisy, hence, the noise $S_{tot}$ at terminal \textbf{A} is
calculated via the superposition principle~\cite{r19},
$S_{total}=t^2_2 \cdot S_1+S_2$, with $S_1(S_2)$  the noise of
QPC1(2). The noise $S_2$ - the own contribution of QPC2 -
indicates whether partitioning events at QPC2 are correlated or
independent.  We find $S_2$ by measuring $S_{tot}$, $S_1$, and
$t_2$ as function of voltage, and use:

\begin{equation}\label{Eq2}
S_2(V_{DS})=S_{tot}(V_{DS})-\int _ {V=0} ^ {V_{DS}} t^2 _2(V)
\cdot {dS_1(V) \over dV} \cdot dV
\end{equation}

\noindent with the integral accounting for the dependence of $t_2$
on $V_{DS}$.  Two examples of the noise produced by a dilute beam,
with average occupation $t_1 \sim 0.2$, are shown in Fig. 3. One
for the dilute beam impinged on a relatively open QPC2 ($t_2 \sim
0.8$) and one when QPC2 is rather pinched ($t_2 \sim 0.4$). The
first striking behavior (insets of Fig. 3) is the apparent
linearity, namely, the weak dependence of $t_2$ on the voltage, in
contrast with the behavior seen in Fig. 1c. Moreover, the
resultant noise $S_2$ (Figs. 3a and 3b) is classical even at the
lowest temperature. Yet, contrary to the results of Comforti et
al.~\cite{r16}, who observed quasiparticles tunnelling through
opaque barriers, here, at a significantly lower temperature,
dilute quasiparticles tend to bunch at strong backscatters.

\begin{figure}
\begin{center}
\leavevmode \epsfxsize=8.5 cm  \epsfbox{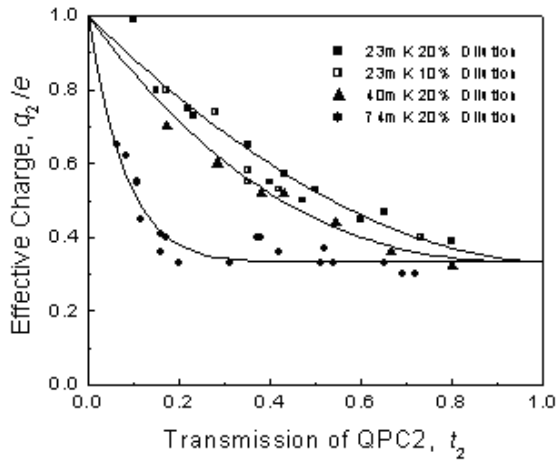}
\end{center}
\vspace{-0.5 cm} \caption{Shot noise resulting from a diluted beam
(with average occupancy $t_1 \sim 0.2$) impinges on QPC2 with
transmission $t_2$. Noise generated by QPC1, $S_1$, in squares and
the total noise measured at A, $S_{total}$, in circles. The
deduced noise generated by QPC2, $S_2$, is plotted in triangles
and is compared with shot noise expected for a binomial
partitioning of quasiparticles with charge $q_2$ (plotted in solid
line). Insets: The differential transmission of QPC2, $t_2$, of an
impinging beam of dilute quasiparticles.  The transmissions are
only slightly energy dependent.  \textbf{(a)}  $S_2$ generated by
a relatively open QPC2, $t_2 \sim 0.8$, and the deduced effective
charge $q_2=0.39e$.  \textbf{(b)} $S_2$ generated by a relatively
pinched off QPC2, $t_2 \sim 0.4$, and the deduced effective charge
$q_2=0.54e$.}
\end{figure}

\noindent With a classical behavior of noise $S_2$ at low
temperatures (Fig. 3), one can determine an effective scattered
charge across a wide temperature range.  The expected $S_2$ at
zero temperature, due to partitioning of charges $q_2$  can be
written as:

\begin{equation}\label{EQ3}
S_2=2q_2 I \cdot t_1 t_2(1-\tilde{t}_2)
\end{equation}

\noindent with $I$ the injected DC current at terminal \textbf{S},
$\tilde{t}_2=t_2[ (e/3)/q_2]$ is the transmission of the particle
\textit{flux} (rather than particle current), each particle with
charge $q_2$~\cite{r15,r16}.  Generalizing Eq. 2 to finite
temperatures allowed to extract an effective charge as function of
transmission $t_2$ at different temperatures. Apparently, the
temperature plays a significant role in the determination of the
effective charge $q_2$ (Fig. 4), while the dependence on small
occupations is weak (similar results for 0.1 and 0.2 occupations).
The charge seemed to increase monotonically as $t_2$ decreased,
however, as the temperature increased the effective charge was
always smaller. This clearly shows that quasiparticle tunnelling
through opaque barriers~\cite{r16} is due to the weaker
correlation among quasiparticles resulting from higher
temperatures.

Note that recently Kane and Fisher predicted~\cite{r17} that at
zero temperature and for an infinitesimal occupation of the
impinging quasiparticles only \textit{electrons} will tunnel even
if the QPC is highly transparent.  We didn't observe such an
effect yet, however, recall that our temperature and occupation
are finite and might not fall in the calculated range of
parameters. We studied here correlation among fractionally charged
quasiparticles that scattered off an artificial impurity in the
FQHE regime ($\nu =1/3$). Adding a new parameter, the occupation
of the impinging quasiparticles, enabled the differentiation
between independent particle-like behavior and condensation of the
scattered quasiparticles to a highly correlated phase, a Chiral
Luttinger Liquid.  Moreover, we find a strong similarity between
diluting the quasiparticle beam and increasing the temperature -
both reduce the particle-particle interaction, rendering the
quasiparticles independent.  Hence, dilution can be employed as a
powerful tool to affect interaction while keeping the ground state
of the system at \textit{zero temperature}.

\textit{Acknowledgements} We wish to thank Yang Ji for his help
during experiment. The work was partly supported by the Israeli
Academy of Science, the German-Israel Foundation(GIF) and the
German Israeli Project Cooperation(DIP).

\end{multicols}
\end{document}